\documentclass[12pt,preprint]{aastex}

\slugcomment{}

\shorttitle{Far-Infrared OH in Sagittarius B2}
\shortauthors{Goicoechea \& Cernicharo}

\begin{document}

\title{Far-Infrared OH fluorescent emission in Sagittarius B2\altaffilmark{1}}

\author{Javier R. Goicoechea and Jos\'e Cernicharo\altaffilmark{2}}
\affil{Instituto de Estructura de la Materia. Departamento F\'{\i}sica 
  Molecular, CSIC, Serrano 121, E--28006 Madrid, Spain.} 
\email{javier@isis.iem.csic.es}
\email{cerni@astro.iem.csic.es}

\altaffiltext{1}{Based on observations with ISO, 
an ESA project with instruments funded by ESA Member States 
(especially the PI countries: France, Germany, the Netherlands 
and the United Kingdom) and with participation of ISAS and NASA.}

\altaffiltext{2}{Visiting Scientist, Division of Physics, Mathematics and 
  Astronomy, California Institute of Technology, MS 320-47, Pasadena, CA 91125, USA}

\begin{abstract}
    
    We present Infrared Space Observatory (ISO)
observations of $^{16}$OH and $^{18}$OH toward
Sgr B2 with a spectral resolution of $\sim$35  km s$^{-1}$.
The OH J=5/2$\rightarrow$3/2 and J=3/2$\rightarrow$1/2 rotational lines of
the $^2\Pi_{1/2}$ ladder are
seen in emission while the cross-ladder transitions
(from the   $^2\Pi_{3/2}$ J=3/2  to the J=1/2, 3/2 and 5/2 levels
of the $^2\Pi_{1/2}$ ladder),
and the   $^2\Pi_{3/2}$ J=5/2$\leftarrow$3/2 and J=7/2$\leftarrow$5/2 lines
are detected in absorption.
The $^{18}$OH  $^2\Pi_{3/2}$ J=5/2$\leftarrow$3/2 $\Lambda$--doublet at $\sim$120 $\mu$m is
also observed in absorption.
All  OH $\Lambda$--doublets are resolved (except
the $\sim$98 $\mu$m) and show, in addition to the strong
absorption at the velocity of Sgr B2, 
several velocity components associated to
the gas surrounding Sgr B2 and to the foreground clouds along the
line of sight.

No asymmetries in the line intensities of each doublet
have been observed.
We have modeled the observations using a non-local radiative
transfer code and found that the OH absorption/emission must arise
in a shell around Sgr B2 not resolved by the ISO/LWS beam. The gas
density is moderate, with upper limits of 10$^4$ cm$^{-3}$
and $\simeq$300 K in temperature. The  OH abundance is high,
(2--5)$\times$10$^{-6}$. 
We argue that a widespread photon dominated region 
explains the enhancement of OH abundance.

\end{abstract}


\section{Introduction}

The Sagittarius  B2 complex represents an interesting  burst of massive star
formation in the Galactic Center (GC) and may be representative of other
active galactic nuclei.
Large scale continuum maps from radio to far--IR wavelengths show that Sgr B2 
is the brightest and among the most massive clouds of the GC (Cox \& Laureijs 
1989, Pierce--Price et al. 2000; Scoville, Solomon \& Penzias 1975).
The Sgr B2 M source is the brightest far--IR condensation of the  complex with
a diameter of $\simeq$40$^{\prime\prime}$ at 100 $\mu$m (Goldsmith et al. 1992) and
has also the largest  gas column density (Lis \& Goldsmith 1989).
These studies have shown that the 
core is embedded in an extended and clumpy cloud of warm gas.
However, the observed rich chemistry in the gas in front of Sgr B2 and
its possible heating  mechanisms are far from  settled. Low velocity shocks 
have been invoked to explain the enhanced abundances of some species such
as NH$_2$ or NH$_3$ (Flower, Pineau des For\^{e}ts \& Walmsley 1995)
which are not observed in more quiescent regions.
On the other hand, UV radiation can have an important effect on the gas
in the outer layers. This radiation may be provided by evolved stars
and by young massive stars in the envelope of Sgr B2 itself 
(Mart\'{\i}n--Pintado et al. 1999).
In addition, the emission of several
ions with high excitation potential (such as [OIII] and [NIII]) is
extended in the Sgr B2 region (Goicoechea et al. 2003, in preparation). Thus,
a widespread ionized component producing photodissociation regions
(PDRs) in the envelope is also possible. 
In any of these scenarios or in a combination of both, 
O--bearing species such as H$_2$O, OH, H$_3$O$^+$ and atomic oxygen
are decisive for the thermal balance of the dense molecular gas 
(Neufeld et al. 1995). In particular, the hydroxyl radical, OH, has
been predicted to be abundant in both scenarios.

The structure of the OH levels is depicted in Figure 1.
The fundamental rotational $\Lambda$--doublet ($\sim$119 $\mu$m) was first
detected in the direction of Sgr B2 in absorption against the thermal dust
emission by Storey, Watson \& Townes (1981). Each line of the doublet is so 
optically thick that absorbs completely the continuum radiation avoiding any
check with the models. These lines are detected over a large area
(9$^\prime\times$27$^\prime$) around Sgr B2 (Cernicharo et al. 1999).
A detailed study of the far-IR spectrum of OH has been carried
out mainly in the Orion shocked region (Watson et al. 1985; Viscuso et al.
1985; Melnick et al. 1987, 1990; Betz \& Boreiko 1989) but little
is known toward GC clouds such as the gas in the line 
of sight toward Sgr B2.

From the theoretical point of view, Offer \& van Dishoeck (1992; hereafter
OfD) and Offer, van Hemert \& van Dishoeck (1994; hereafter OfHD) have made
detailed calculations of the collisional cross sections between OH and H$_2$
and of the expected intensity of the OH far--IR lines. They predicted 
strong asymmetries in the intensity of the OH $\Lambda$--doublets due
to important asymmetries in the collisional rates between OH and para-H$_2$.

In this letter, we present high--resolution observations of several
OH $\Lambda$--doublets toward Sgr B2 involving levels up to $\sim$420 K.
All  lines inside
the $^2\Pi_{1/2}$ ladder are seen in emission while all the remaining lines
are detected in absorption. The data have been modeled using a non-local 
radiative transfer code in order to constrain some important physical 
parameters of the warm outer layers of Sgr B2 and of the foreground gas.

\section{Observations, Data Reduction and Results}

Most of the pure rotational lines  of OH are in the far--IR wavelength
coverage of the Long-Wavelength Spectrometer
(LWS; Clegg et al. 1996) on board ISO (Kessler et al. 1996).
The ISO data base observations used in this Letter are
32201429,
32600502,
46201118
46201123,
46701803,
47600809,
47608099,
49401705,
50400823,
50601013,
50600603,
50700610, 
and 84500102.
The LWS/Fabry--Perot (FP) spectral resolution  is $\simeq$0.015 $\mu$m
($\sim$35  km s$^{-1}$) and allows resolving the $\Lambda$--doubling
of OH and its isotopes.
Several OH rotational lines
(intra and cross--ladder) were searched toward Sgr B2 M. Data are
shown in Figure 1.
The OH J= 5/2$\leftarrow$3/2  cross-ladder doublet at $\sim$34 $\mu$m was
observed with the Short-Wavelength Spectrometer (SWS; de Graauw et al. 1996) with a
spectral resolution of 1500 (ISO observation 28702002). 
One component  was observed with the SWS/FP with a velocity resolution of 
$\sim$10 km s$^{-1}$ (ISO observation 46001217).
When compared to the grating observation, the
mean LWS/FP continuum flux of each line deviates by at least 20\%, similar to
the grating flux calibration errors (Swinyard et al. 1998). The 
systematic uncertainties are always larger than the statistical errors
associated with the line signal--to--noise ratios (S/N). The best  
calibration is obtained by correcting the continuum level of the FP
spectrum 
in order to coincide with the grating level.
A polynomial baseline was fitted to each spectra and adopted as the
continuum level.
The effective SWS/FP aperture in the $\sim$34 $\mu$m range
is 17$^{\prime\prime}$$\times$40$^{\prime\prime}$,
while the LWS aperture is circular and $\sim$80$^{\prime\prime}$ in diameter.
All the SWS and LWS products were  processed with
version 10.1 of the pipeline and analyzed using the ISO spectrometers data 
reduction package ISAP.\footnote{The ISO Spectral Analysis Package (ISAP) 
 is a joint
development by the LWS and SWS Instruments Teams and Data Centers. 
Contributing institutes are CESR, IAS, IPAC,MPE, RAL, and SRON.}

Sgr B2 M has been the target of several spectroscopic observations
in the far--IR. Molecular features have been always found in absorption. 
However, OH shows a different behavior. All rotational doublets connecting
the $^{16}$OH  $^2\Pi_{3/2}$ ground--state with the
rotational levels of the $^2\Pi_{1/2}$ ladder ($\sim$34, 53, 79 and 119
$\mu$m) are observed in absorption while the $^{16}$OH $^2\Pi_{1/2}$ 
intra-ladder rotational transitions at $\sim$98 and $\sim$163 $\mu$m are 
observed in emission. 

The $\sim$79 and $\sim$53 $\mu$m $\Lambda$--doublets have maximum absorption 
depths at the velocity of Sgr B2 M (V$_{LSR}$$\simeq$+50  km s$^{-1}$).
Nevertheless, there is
still a  contribution from more negative velocity components associated
with the ``spiral arms clouds'' in the line of sight and with the gas
surrounding the GC, which we divide in two components ($\simeq-85$
and 0 km s$^{-1}$). The n(H$_2$) range found in these clouds varies
from 10$^2$ to 10$^4$ cm$^{-3}$ (Greaves 1995).
The absorption depths at V$_{LSR}$$\simeq$-85 km $s^{-1}$ are
$\simeq$30$\%$  and $\simeq$15$\%$ 
for the $\sim$79  and $\sim$53 $\mu$m doublets respectively.
These components  are thought
to arise in the molecular gas within 1 kpc of the GC (Scoville 1972).
Assuming that the absorption at these velocities 
is optically thin and considering only radiative
excitation (i.e., low T$_{ex}$),
we derive $\chi(OH)\sim10^{-6}$ for
$N(H_2)$$\simeq$(5--10)$\times$10$^{21}$ cm$^{-2}$ (see Neufeld et al.
2000 for a distribution of $N(H_2)$ in the line of sight). The
opacity of the OH lines at V$_{LSR}$$\simeq$0  km s$^{-1}$
allows also to derive $\chi(OH)\sim5\times10^{-7}$ for
$N(H_2)$$\simeq$2.5$\times$10$^{22}$   
cm$^{-2}$. This component is believed to arise in the diffuse gas at 
galactocentric radii 3--8 kpc and in the Solar vicinity
(Greaves \& Williams 1994).
 These values are higher than the typical
OH abundances of 10$^{-7}$ found in diffuse and translucent clouds
(Van Dishoeck \& Black 1986).

The $\sim$84 $\mu$m doublet and the  34.603 $\mu$m line component (in
absorption) and the $^2\Pi_{1/2}$ doublets at $\sim$98 and
$\sim$163 $\mu$m (in emission) are  strictly centered at
the velocity of Sgr B2 M  and are decisive to understand the excitation
scheme of the molecule.
The ISO data also provide upper limits to the emission/absorption
of the J=3/2$\rightarrow$5/2 ($\sim$96 $\mu$m) and J=5/2$\rightarrow$5/2
($\sim$48 $\mu$m) cross--ladder
transitions
and to the $^2\Pi_{3/2}$ J=9/2$\rightarrow$7/2 ($\sim$65 $\mu$m) and 
$^2\Pi_{1/2}$ J=7/2$\rightarrow$5/2 ($\sim$71 $\mu$m)
$\Lambda$--doublets  of 1.5\% of the continuum emission.
Finally, we present the detection of the fundamental $^{18}$OH $^2\Pi_{3/2}$
J= 5/2$\leftarrow$3/2 lines at $\sim$120 $\mu$m.
Apart from the feature associated with Sgr B2, there is another absorption 
at V$_{LSR}$ $\simeq$0  km s$^{-1}$. The derived $^{18}$OH 
column density is (2.5$\pm$0.5)$\times$10$^{13}$ cm$^{-2}$, implying
an $^{16}$OH/$^{18}$OH abundance ratio in this component of $\sim$500,
similar to the terrestrial value.

\section{Modeling Sgr B2 and Discussion}

The typical OH abundance in dense quiescent clouds is
(0.1-1)$\times$ 10$^{-7}$ with OH/H$_2$O ratios in the range
1 to 10$^{-2}$ (Bergin et al. 1995).
According to the chemical models, the major contribution to the
enhanced  OH abundance 
comes from regions where water vapor is being rapidly photodissociated
(Sternberg \& Dalgarno 1995) or regions in which shock waves play a role 
(Draine et al. 1983). 
In the case of Sgr B2, the innermost regions of the cloud
are completely hidden in the mid and far-infrared due to the large dust
opacity. Almost all the observed far--IR OH  come from the
external layers of the Sgr B2 cloud, i.e., from the surrounding gas at 
high T$_K$ and from the
cold dark clouds and diffuse gas in the line of sight.

In order to estimate the OH abundance and the physical conditions leading
to the observed OH line emission in the V$_{LSR}\simeq+50$  km s$^{-1}$
component, we have modeled the first 20 OH rotational
levels using the code developed by Gonz\'alez-Alfonso \& Cernicharo
(1993). The hyperfine structure of OH is not included in the model.
The collisional cross sections of OfHD have been used 
(kindly provided in electronic form by E.F van Dishoeck).
We have adopted a spherical geometry for a cloud consisting of two components:
a uniform core with a diameter of 25$^{\prime\prime}$ (3.2$\times$10$^{18}$ cm for
an assumed distance to Sgr B2 of 8.5 kpc), with a dust
temperature of 30 K (see Goicoechea \& Cernicharo, 2001)
and opacity at 80 $\mu$m of 5
($\tau_{\lambda}=\tau_{80}*(80/\lambda({\mu}m)$),
and a shell of
variable thickness and distance to the core. The shell was divided
into 14 layers. All molecular transitions in the core are thermalized to
the dust temperature due to the large dust opacity in the far--IR.
In order to check the sensitivity of the results on the physical 
parameters, 
$n(H_2)$, T$_K$ and $N(OH)$
were varied from 10$^3$ to 10$^5$ cm$^{-3}$, 40 to 600 K, and
1$\times$10$^{15}$ up to 1$\times$10$^{17}$ cm$^{-2}$ respectively.

The observations put some constraints on the size and density of the
shell. Excitation temperatures for the cross-ladder and $^2\Pi_{3/2}$ 
transitions have to be below the dust temperature in order to reproduce
the observed line absorptions. In addition, the shell thickness 
and its distance to the core cannot be large compared to the
core size, otherwise limb effects introduce important
re-emission that could cancel the absorption produced by the gas in
front of the core. The results for some  models are reproduced in
Figure 2. We have found that if the shell is placed at some distance 
from the core, then it is difficult to reproduce the observations as
almost all lines appear in emission within the LWS beam
(80$^{\prime\prime}$). Even if the shell is contiguous to the core but
large in thickness, lines  also appear in emission. 
Models in Fig. 2 correspond to a total size of 42$^{\prime\prime}$
(5.3$\times$10$^{18}$ cm) for the core+shell cloud. 

The model labelled M$_1$
corresponds to T$_K$=40 K, $n(H_2)$=10$^4$ cm$^{-3}$ and 
$\chi(OH)=2\times10^{-6}$. The cross-ladder and $^2\Pi_{3/2}$
transitions are in absorption while the $^2\Pi_{1/2}$ intra-ladder
transitions are predicted in emission.
However, although the intensity of the absorbing lines is found
to fit reasonably well the observations, emission lines are
underestimated. If the OH abundance  is increased in order
to reproduce the intensity of the emitting lines then, 
absorbing lines are poorly fitted as limb effects start to
dominate. This effect is already  seen in the J=1/2$\leftarrow$3/2
 cross-ladder transition at $\sim$79 $\mu$m. In particular, the $^2\Pi_{1/2}$
 J=5/2$\rightarrow$3/2  transition at $\sim$98 $\mu$m
is strongly underestimated by model M$_1$. If the gas temperature 
increases, collisions start to pump the OH levels and the intensity of the
$\sim$98 $\mu$m  line is better fitted. 
However, a new problem arises as strong
asymmetries in the line intensities of each $\Lambda$-doubling line
do appear at high temperature  (no asymmetries have been observed within the 
S/N ratio of the data ). These asymmetries were already predicted
by OfD and are due to a strong parity change propensity rule introduced
by collisions with para--H$_2$.
They can only be suppressed if the H$_2$ ortho--to--para (OTP) ratio is higher or if
radiative pumping dominates.
We have adopted a OTP ratio of 3.
Model M$_2$ in Fig. 2 corresponds to
T$_K$=150 K with otherwise the same parameters
as M$_1$. The intensity of the  $\sim$98 $\mu$m doublet
is considerably enhanced with respect to the low temperature
model. The intensity asymmetry in this doublet is not visible
in the data as the doublet is unresolved with the LWS/FP
spectrometer.
The effect of the IR pumping strongly reduces the asymmetries
that could be present without the optically thick core.
Finally, model M$_3$ (thick lines in Figure  2) represents 
T$_K$=300 K, $n(H_2)$=5$\times$10$^3$ cm$^{-3}$  and
$\chi(OH)=3\times10^{-6}$. Again, the strong asymmetry of the  $\sim$98 $\mu$m
doublet is not  observable due to the lack of spectral
resolution. However, the observed total intensity for the doublet
agrees well with that predicted from the model. For the other lines, M$_3$
reproduces qualitatively well the observed OH pattern, except
in the line wings in the 119 and 79 $\mu$m
doublets. Nevertheless,
cold foreground gas (not included in the model) absorbs
(like in M$_1$) at these
wavelengths, reducing the  emission wing effects in these
lines.

From our models it is clear that the gas surrounding Sgr B2 has to be
warm (see Ceccarelli et al. 2002) and must
have a moderate H$_2$ density. The OH abundance is rather high, 
(2--5)$\times$10$^{-6}$. Assuming that M$_3$ represents a reasonable
fit to the OH data,
we derive $N($$^{18}$$OH)$=(6$\pm$2)$\times$10$^{13}$ cm$^{-2}$ for Sgr B2. 
Hence, the $^{16}$O/$^{18}$O isotopic ratio is 240--280, in excellent agreement
with the value obtained
by Bujarrabal, Cernicharo \& Gu\'elin (1983) from the OH 18 cm lines and 
a factor $\simeq$2 lower than in the V$_{LSR}$ $\simeq$0  km s$^{-1}$ component.
Higher T$_K$ could even be possible but the actual collisional
rates  introduce very strong intensity asymmetries in the
cross-ladder transitions which are not observed. If the gas temperature 
is higher, then its density has to  be lower than the value used
in M$_3$ (5$\times$10$^3$ cm$^{-3}$).
The reason is that higher temperatures produce strong emission
in the OH high excitation transitions 
($\sim$71 and $\sim$65 $\mu$m for example, see Figure 2)
which is not observed. For T$_K$=600 K, $n(H_2)$ has to be
decreased to (1-2)$\times$10$^3$ cm$^{-3}$ and $\chi(OH)$ increased to
(0.5-1)$\times$10$^{-5}$ to obtain a crude fit the ISO data. A temperature
gradient across the shell, going from T$_K$=40 K in the innermost
regions to T$_K$=600 K in the external layers is probably a better
representation to the physical structure of the edge of the Sgr B2 cloud.

The inferred $^{16}$OH column density in the shell, 
(1.5--2.5)$\times$10$^{16}$ cm$^{-2}$, and that of H$_2^{16}$O 
determined also from LWS/FP data (Cernicharo et al. 1997; 2002, in preparation)
leads to an OH/H$_2$O=0.1--1 abundance ratio.
For comparison, shock models of Flower et al. predict $\simeq$$10^{-3}$,
while models of dense dark cores predict $\simeq$10$^{-4}$.
On the other hand, the maximun OH abundance in PDR models is $\simeq$10$^{-5}$
with OH/H$_2$O ratios close to 10.
The large abundance found in Sgr B2 suggests that its external shells
are illuminated by a strong UV field producing a PDR, although low
velocity shocks (v$_S\sim$30 km $s^{-1}$),
 unresolved by the ISO data,  may be also present. 
Many species formed during the evolution of the cold gas in Sgr B2 are
being now reprocessed in these regions. This warm gas is poorly traced by
millimeter and submillimeter observations, but it represents the strongest 
contribution to the absorption/emission features in the far--IR spectrum
of Sgr B2. In addition to $^{16}$OH and $^{18}$OH, we have searched for several 
related species such as  OH$^+$ or H$_2$O$^+$. No lines from these
species have been found.

The future far--IR instruments
on board next generation telescopes such as the 
{\it{Herschel Space Observatory}} will allow a fast mapping  of many OH 
rotational lines in galactic and
extragalactic sources making OH a very useful tool in 
deriving the physical properties of 
GC giant molecular clouds. This could help in the analysis of starburst
galaxies like Arp 220.

\acknowledgments
We thank Spanish DGES and PNIE for funding
support under grant ESP98--1351--E and PANAYA2000-1784
and  C.M. Walmsley, V. Bujarrabal and J. Fischer for their comments on
the manuscript. We also thank our referee, E.F. van Dishoeck, for her
useful comments and suggestions.
JRG acknowledges \textit{UAM} for a pre--doctoral fellowship.

\clearpage

\begin{figure}
\caption
{
  ISO observations of $^{16}$OH and $^{18}$OH
  toward Sgr B2 M and rotational energy diagram of OH. The two ladders,
  $^2\Pi_{3/2}$ and $^2\Pi_{1/2}$, produced by the spin orbit interaction
  are shown. The $\Lambda$--doubling splitting of each rotational
  level has been enhanced for clarity. The hyperfine structure of the
  rotational levels is not shown. The observed rotational transitions
  are indicated by arrows and the observed ISO line profiles are 
  shown in boxes.
  The intensity scale
  corresponds to the continuum normalized flux and the abscissa to the
  wavelength in microns. When lines are centered only at Sgr B2 velocities
  ($\simeq$+50 km s$^{-1}$), the LSR velocity scale is not shown.
  All $\Lambda$--doublets are LWS/FP data excepting the 34.6 $\mu$m
  doublet for which  SWS grating data ($\lambda/\Delta\lambda\simeq$1500)
  are shown. One component has been observed with the SWS/FP
  ($\lambda/\Delta\lambda\simeq$30000) and data are shown in the inset of
  the corresponding box. The $^{18}$OH  $^2\Pi_{3/2}$ J=5/2$\leftarrow$3/2
  doublet is shown in the upper inset.}
\end{figure}

\begin{figure}
\plotone{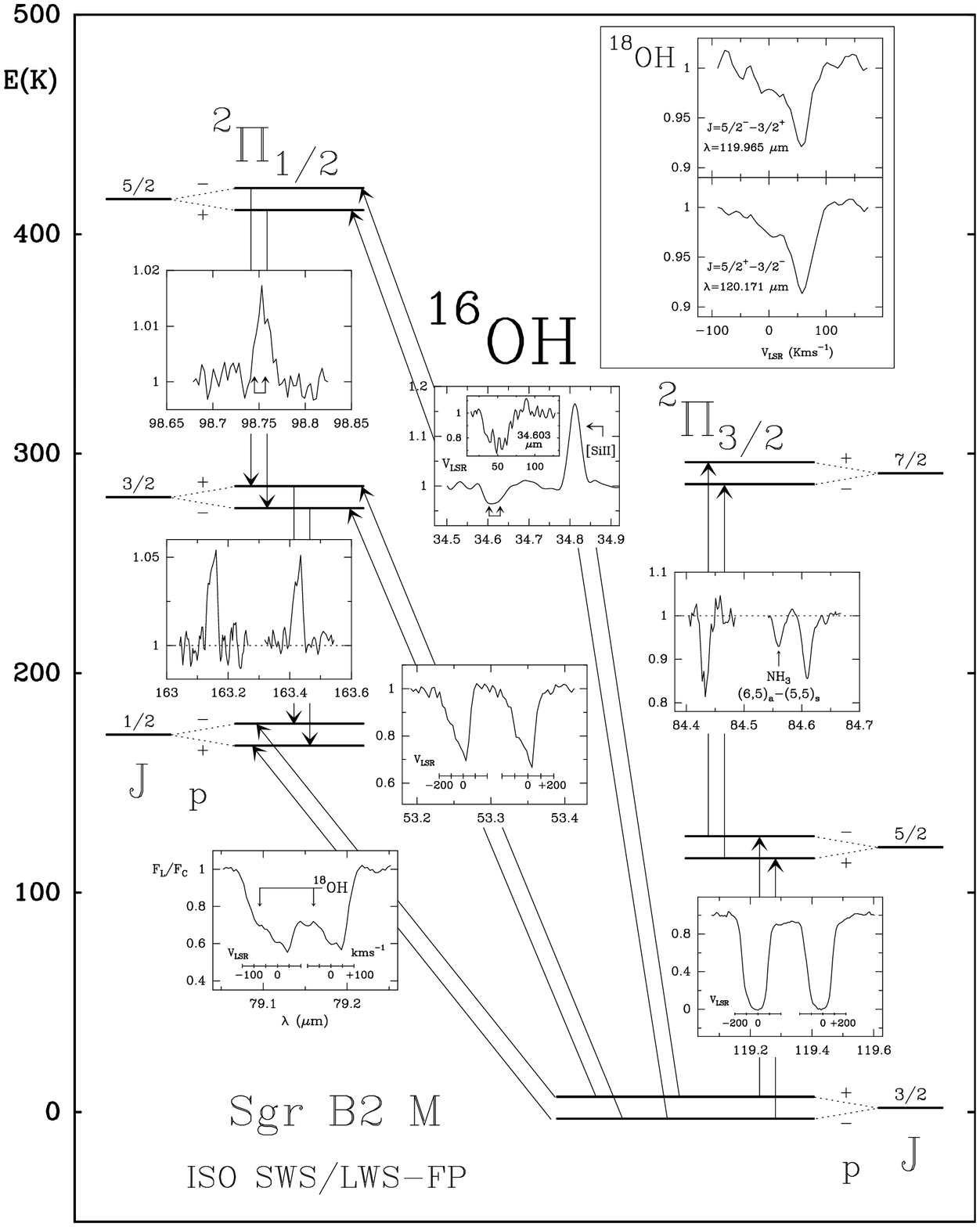}
\end{figure}

\clearpage 

\begin{figure}
\caption
{
  Results from selected models of Sgr B2 component (V$_{LSR}\simeq+50$  km s$^{-1}$)
  discussed in the text. The central panels
  correspond to the cross-ladder transitions
  of OH. Line intensities have been normalized to the continuum dust
  emission. The wavelength of the transitions is indicated at the
  right of each panel. The  model labelled M$_1$ corresponds to T$_k$=40 K,
  n(H$_2$)=10$^4$ cm$^{-2}$, $\chi_{OH}$=2$\times$10$^{-6}$.
    The model labelled M$_2$ corresponds to T$_k$=150 K,
  n(H$_2$)=10$^4$ cm$^{-2}$, $\chi_{OH}$=2$\times$10$^{-6}$.
  The  model labelled M$_3$ corresponds to T$_k$=300 K,
  n(H$_2$)=5$\times$10$^3$ cm$^{-2}$, $\chi_{OH}$=3$\times$10$^{-6}$.
  The velocity resolution in the modeled line profiles is 1 km
  s$^{-1}$.}
\end{figure}

\begin{figure}
\plotone{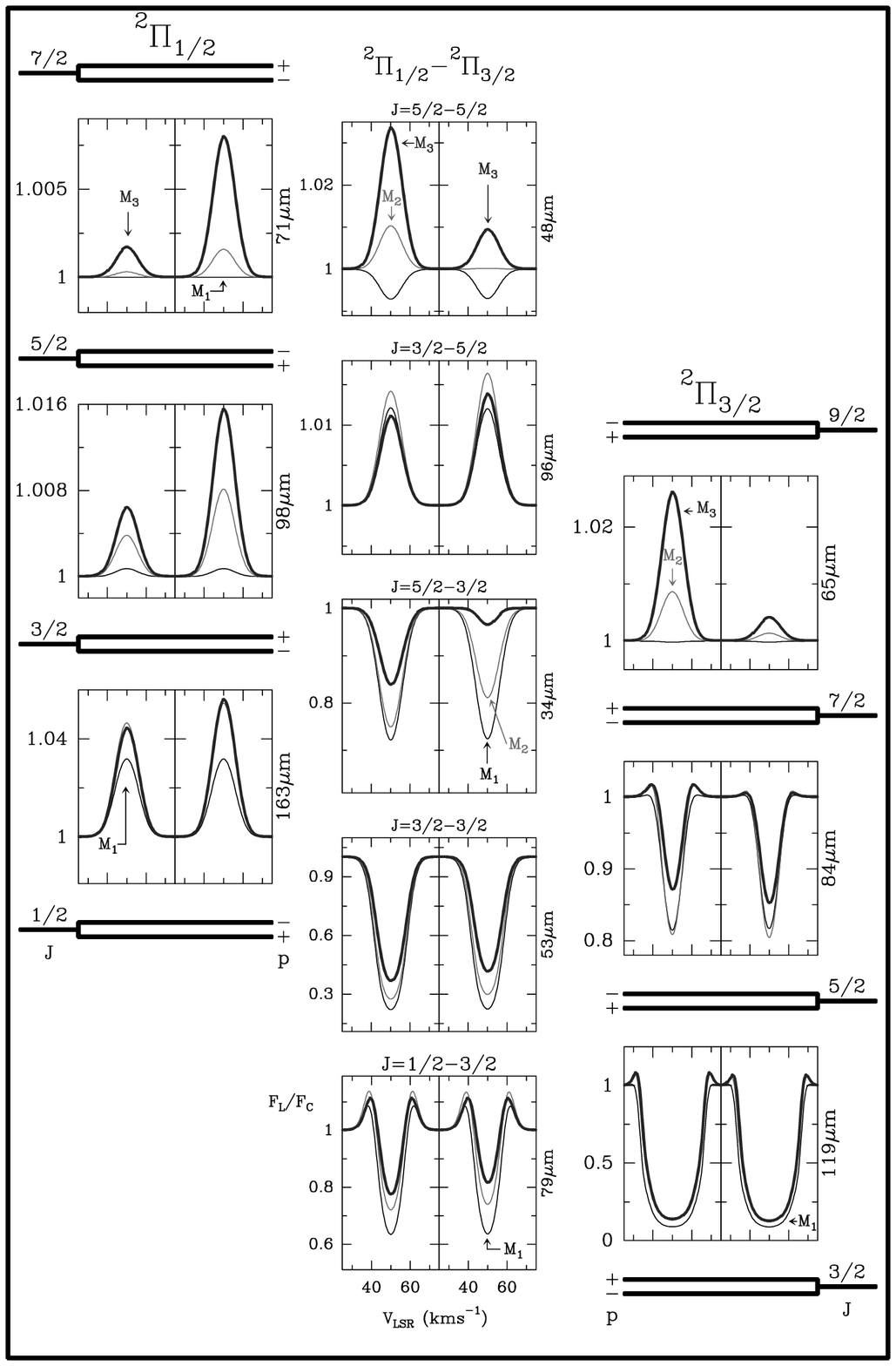}
\end{figure}

\end{document}